\documentclass[10pt,a4paper,twoside]{article}
\usepackage{epsfig}
\usepackage{baltlat5}
\usepackage{wrapfig}
\def\shah{{\rm III}}

\begin{document}
\ \
\vspace{-0.5mm}

\setcounter{page}{1}
\vspace{-2mm}

\titlehead{Baltic Astronomy, vol.\ts xx, xxx--xxx, 2009.}

\titleb{HIGH FREQUENCY LIMITS IN PERIODICITY SEARCH FROM IRREGULARLY SPACED DATA}

\begin{authorl}
\authorb{J.~Pelt}{}
\end{authorl}

\moveright-3.2mm
\vbox{
\begin{addressl}
\addressb{}{Tartu Observatory, 61602 T\~{o}ravere, Estonia}
\end{addressl}
}

\submitb{Received 2009 February 12; revised 2009 April 24}

\begin{summary}
Notions and limits from standard time series analysis must
be modified when treating series which are measured irregularly and contain
long gaps. Classical Nyquist criterion to estimate frequency range
which is potentially recoverable must be modified to handle
this more complex situation. When basic exposition of the modified criterion
is given in earlier papers, some minor problems and caveats are treated here.
Using simple combinatorial arguments we
show that for small sample sizes the modified Nyquist limit may overestimate
the obtainable frequency range.  On the other hand we will demonstrate that
very high Nyquist limit values which are typical to irregularly sampled data
can often be taken seriously and using proper observational
techniques the frequency ranges for ``time spectroscopy'' can be significantly
widened.
\end{summary}


\resthead{High frequency limits in periodicity search}{J.~Pelt}

\sectionb{1}{INTRODUCTION}
It is very often the case that observed time series seems to be variable but
physical origin of the variability is unknown. The first thing what astronomer
does in this case is to compute standard power spectrum or similar statistic for a wide range of
trial frequencies and look for possible peaks. But how far along frequency axis one should go with this
analysis? Typical answers can be found e.g. in Kurtz~(1983), Eyer \& Bartholdi~(1999) and
Koen~(2006).
Below we look at this question from different angles trying to clarify some theoretical aspects which are untreated till now. Some of
our observations show that proposed schemes to estimate upper reasonable frequency limits
can be misleadingly high. However, in most cases one tends to underestimate the detection
potential of precisely timed and truly randomly spaced measurement series.

\sectionb{2}{EQUALLY SPACED DATA}

In traditional Fourier analysis the highest frequency that can be extracted from
equally spaced and continuously monitored data is $s_{Nyq}={1\over 2\Delta t}$ (where
$\Delta t$ is sampling step in time) - so called {\it Nyquist frequency} (in the context of astronomy, see e.g. Kurtz~1983). 
This fact can be proved using different analytical tools. For our purposes the shortest route to
the result comes from observation that regularly spaced sampling can be looked upon as
a multiplication of continuous input function with appropriately scaled in time
$\shah$
function (see Bracewell~2000):
\[
\shah(t) = \sum\limits_{n =  - \infty }^\infty  {\delta (t - n)}.
\]
To the multiplication in time domain corresponds convolution in Fourier domain.
The Fourier transform of a $\shah$ function is also a $\shah$ function. As a result, we get a sum of periodically (with period $1\over \Delta t$) repeated original spectra. 
If the frequency spectrum $F(s)$ of original data $f(t)$ is bandlimited by the Nyquist frequency ($F(s)=0 {\rm ~when} \left| s \right| >= \frac{1}{{2\Delta t}}$) then its full recovery is possible. 
Otherwise different shifted replicas of the original spectrum overlap and exact recovery is impossible.

\vbox{
\centerline{\psfig{figure=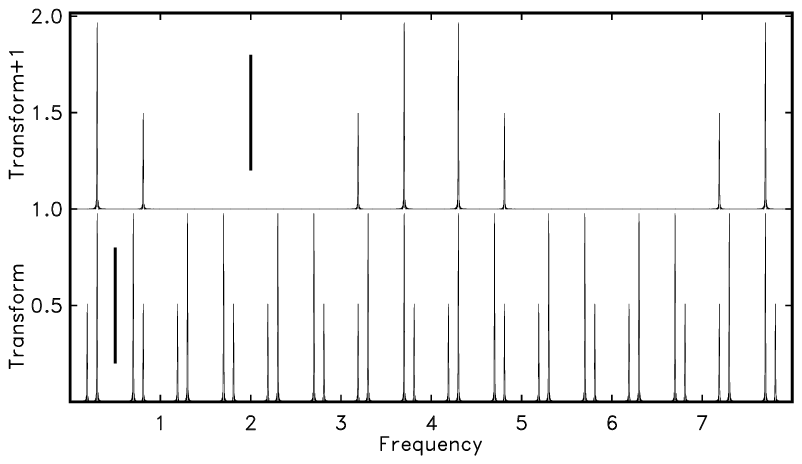,width=120truemm,angle=0,clip=}}
\vspace{-5mm}
\captionc{1}{Fourier transforms of $f(t)=0.5\cos(2\pi t \times 0.81)+\cos(2\pi t \times 0.3)$. For lower spectrum $\Delta t = 1$, for upper spectrum $\Delta t= 0.25$. Nyquist limits are
depicted using thick lines.}
}
\vspace{3mm}

This is well illustrated on Figure 1 where two Fourier transforms of the input signal $f(t)=0.5\cos(2\pi t \times 0.81)+\cos(2\pi \times 0.3)$ are depicted. In upper panel
the sampling step $\Delta t = 0.25$ is short enough to reproduce both frequencies correctly. Spurious peaks start to occur only for frequencies which are higher than
Nyquist frequency (in this particular case $s_{Nyq}=2$). In lower panel the transform is computed from data with time step $\Delta t = 1$. Now the replicas of the
original spectra are spaced more densely and inside of the Nyquist limits we can see correct peak at frequency $s=0.3$ and spurious peak at frequency $s=-0.81+1=0.19$. This kind of
overlapping of shifted replicas of the spectrum is generally known as aliasing.

Very often aliasing effect is misinterpreted in a certain way. Normally it is assumed that
all non-zero energy of oscillations is situated in the first replica around zero frequency. But this need not be the case. If the actual frequency of the input process is
much higher than the Nyquist frequency then its aliases can show up at lower frequencies. From the convolution
theorem follows only periodic replication of the spectral fragments, not the actual position
of the true replica. In some cases this observation allows  to observe
high frequency phenomena with apparatus which samples input data at significantly
lower rate from that required by the Nyquist theorem. The physically relevant frequency
from the set of periodically spaced candidates (with step in frequency $1\over \Delta t$) can then be selected using some additional criteria.

\sectionb{3}{IRREGULARLY SPACED DATA}

In the case of data sets where time moments are measured irregularly the Fourier convolution theorem can be similarly applied (see Deeming~1975). 
Instead of the $\shah$ function we can now introduce {\it data window function} $w(t)$:
\[
w(t) = \sum\limits_{n = 1}^N {\delta (t - t_n )},
\]
for time point sequence $t_n, n=1,\dots,N$. The Fourier image of the sampled data
can now be described as a convolution of the real continuous data spectrum $F(s)$ with the Fourier transformed data window function $W(s)$. Typical window function
is depicted on Figure 2.
 
\vbox{
\centerline{\psfig{figure=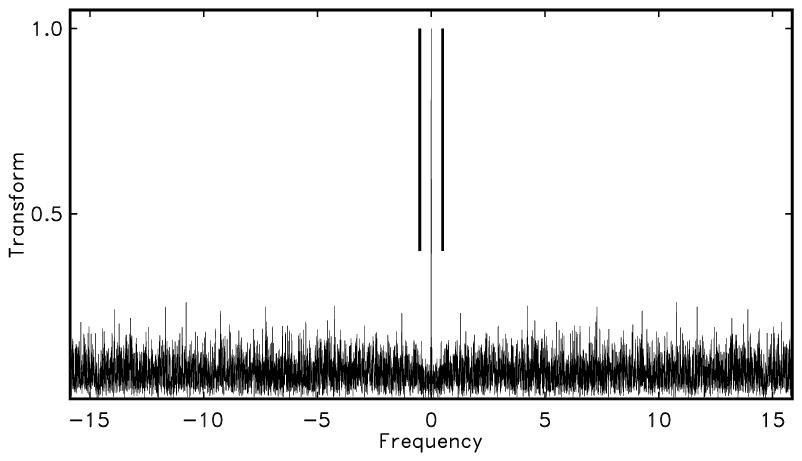,width=120truemm,angle=0,clip=}}
\vspace{-5mm}
\captionc{2}{Typical Fourier transformed data window function $W(s)$. Nyquist limits are depicted using thick lines.}
}
\vspace{3mm}

The strongest peak is at zero frequency as it should be. But instead of periodically occurring side peaks we can now see
only random peaks with significantly lesser amplitude. The important point for a following discussion is that fluctuations inside the Nyquist limits
are not very much different if to compare with fluctuations far away.

Now we have a reasonable question - how to put correct limits to frequency search ranges if our input data is randomly sampled?

There are essentially three methods discussed till now, each one based on
different heuristic deliberations.

First, it is possible to look at irregularly spaced data set as a disturbed
version of the regularly spaced data. Consequently the mean time step is
calculated $\bar \Delta t = (t_N  - t_1 )/(n - 1)$ and corresponding Nyquist limit is calculated using this value. 
This view is supported also by statistical
argument - if we compute spectra from $N$ statistically independent input points 
then there cannot be more than $N/2$ statistically independent spectrum
points. (Another half of the input statistical information is spent on negative frequencies or phases).
However, simple trial calculations show that
so calculated limits are too conservative and useful information can be lost
(see Press et al~2007).

Then it is argued that in the case of irregularly spaced data distances between
some consecutive time points are significantly shorter than mean time step and this allows define the Nyquist frequency using minimal time difference between two consecutive observations.
Again the numerical experiments (see Eyer \& Bartholdi~1999) show that
this is not the case and corresponding criterion is too conservative.

The third method (see Eyer \& Bartholdi~1999, Koen~2006) takes off from the observation that actual times of observation are always of limited precision and
correspondingly their values form a subset of a certain fixed step grid. The ultimate
Nyquist limit then can be computed from time step of this grid. The limit can be made
more sharp by checking phase distributions for the multiples of the minimal step
(check Eq. 7 of the Koen~2006).

\sectionb{4}{RETURN ARGUMENT}

The third method to estimate the Nyquist frequency is certainly valid, 
but it contains one important hidden caveat.

Let the Nyquist frequency computed for a particular data set be $s_{Nyq}$. If we want to cover
in our analysis frequency range $(0,s_{Nyq})$ with proper frequency resolution 
we need check at least $K=s_{Nyq}(t_N-t_1)$ trial frequencies (proper step along frequency axis is $\Delta s = {1\over t_N-t_1}$). 
For a extensive set of time point arguments so computed $K$ can be quite large. It is not
ruled out that the number of essentially different phase configurations exhausts itself
earlier than $K$ trials are computed. 

To illustrate this we look first at one old, but still popular method of period seeking, the
so called Lafler-Kinman~(1965) method. In this method for each trial
frequency $s$ phases
\[
\phi (t_n ,s) = s \times t_n  - [s \times t_n ],n = 1, \ldots ,N
\]
are computed (squared brackets denote taking of integer part). Then data points $m(t_n),n=1,\ldots,N$ are sorted according to the
phases. For sorted values (say, magnitudes) the final frequency dependent
statistic is computed
\[
\theta (s) = \frac{{\sum\limits_i {(m_i  - m_{i + 1} )^2 } }}{A}
\]
where $m_i$-s are just sorted values for increasing phases, summation takes
into account cyclicity ($m_{N+1}=m_1$) and $A$ is certain frequency independent scaling factor. The minima of the $\theta(s)$ function indicate then most probable
frequencies (it is kind of inverted power spectrum). Very often the methods
based on this statistic and similar to that are called as ``string length'' methods.

For us the important point is that actual frequency dependent phases do not enter
into final statistic, their effect is accounted for only through generated
sort orders. But there are only $N!$ different possible orderings (permutations).
Consequently, if there is, say only 5 observations in our data set, then there
is just $120$ different possible permutations and possible $\theta(s)$ values.
It is guaranteed that $\theta(s)$ values start to reoccur and there is no way to distinguish
between them. However for high precision observations the formal Nyquist range can
contain much more trial frequencies! As we see 
the low number of observations themselves puts a certain limit to the number of
frequencies which can be inspected using this kind of statistic. Or to put it in another way around the
phase configurations {\it start to return} before all $K$ trial frequencies are checked through.

Of course, it can be said that Lafler-Kinman statistic is of very peculiar kind
and we just can avoid it. Unfortunately this is not the case. For other statistics
it is quite easy to involve similar arguments.

For instance in methods where statistics are computed using binning of phases
(see for instance Jurkevich~1971 or Schwarzenberg-Czerny~1989) there can be only
$M^N$ essentially different phase configurations (for each from $N$ phases we can 
assign arbitrary bin from $M$ bins available). For small values of $N$ and $M$ it
can then happen that $K>M^N$ and statistic values computed from different phase configurations start to return.

For statistics which depend continuously from phases 
(standard Fourier spectrum, Lomb-Scargle method {\it etc}) situation is more complicated.
In principle for a set of precisely measured incommensurable time points the phase configurations
and computed from them statistic values never return exactly. But if we fix a certain level of
precision it is always possible to estimate approximate return times.   

If we imagine frequency dependent phases as points
on the circle which rotate with incommensurable speeds then this system will return
arbitrary close to whatever state after certain time. This so called Poincar\'e cycle
return time can be approximately estimated (see e.g. Kac~1947) and
in our context it scales as $N^N$. The heuristic argument behind this value is simple.
We can divide the full range of phases $[0,1)$ into $M$ bins and approximate continuous
base functions (say $\cos$ and $\sin$) in piecewise constant manner. This allows us to
involve return argument for binned case described above. To use all relevant phase
information we can select binning scheme with $N$ bins and here we are - approximately equal statistic values
start to return not later than after $N^N$ trials.

From the first sight restriction imposed by phase configuration return argument is relevant 
only for data sets with very low number of observations and need not be taken seriously. 
However, very often we have data sets which contain considerable number of observations but with peculiar 
distribution where the observing moments tend to be concentrated in small number of densely populated groups. Now the return
argument starts to work for a low frequency part of the spectrum. For a periods
which are significantly longer than group lengths the groups behave as singular points.

\break

\sectionb{5}{INTEGRATION TIME EFFECTS}

Real observations of the variable objects are always obtained using certain
integration (or exposure) times to obtain reasonable photon counts and from it a measurable output signal. Let us assume that integration time for every single
observation is $\delta t$. Then the measurement procedure can be formally described as a sequence of two operations: convolution of the input continuous waveform by
rectangle function $\frac{1}{{\delta t}}\Pi \left( {\frac{t}{{\delta t}}} \right)$
followed by the sampling proper. In the time domain to the convolution with a rectangle function  corresponds multiplication with ${\rm sinc}$ function:
\[
{\rm sinc} (s\delta t) = \frac{{\sin (\pi s\delta t)}}{{\pi s\delta t}}
\]
in the Fourier domain. This fact introduces additional restriction to
the reasonable frequency range in periodicity search. The convolution acts
as a low pass filter and higher frequencies are strongly attenuated. Typical
frequency limits obtained from this analysis are in the range
from $s_{\max }  = \frac{1}{{2\delta t}}$ to $s_{\max }  = \frac{1}{{3\delta t}}$,
(see for instance Eyer \& Bartholdi~1999).

Normally the side lobes of the ${\rm sinc}$ functions are ignored and corresponding
higher frequencies are left out from analysis. However, in the case when exposure times
are sufficiently exactly measured and exposures are spaced in time sufficiently randomly
it is possible to find a proper number of observations $N$ for which peaks in frequency space side lobes can be
correctly detected. To cover regions where attenuation is very strong (say around zero crossings) we can use
different exposition times. This methodology needs further elaboration and will be treated elsewhere.

\sectionb{6}{MODIFIED NYQUIST LIMITS CAN BE TAKEN SERIOUSLY}

In previous work (e.g. Press et al~2007, Eyer \& Bertholdi~1999 or Koen~2006) the peculiar alias free nature of
the irregularly measured samples is demonstrated by computing sample spectra for
$2-5$ times wider ranges than the ``normal'' Nyquist ranges (computed from average
time step length). From phase configuration return argument above we can conclude
that for moderately large values of $N$ phase configurations do not return. Consequently
we can safely apply formal Nyquist limits. For precisely measured time point series
these limits can be quite high.

To check this in practice we performed some very simple
and therefore easily repeatable calculations. First we postulated
a standard harmonic model for our data sets:
\[
f(t_n ) = \cos (2\pi st_n ) + E\varepsilon _n ,{\rm{  }}n = 1, \ldots ,N
\]
where $\varepsilon_n$-s are normally distributed random variables with unit dispersion and
$E$ is parameter which controls noise level. The time point sequence $t_n,n=1,\ldots,N$ was randomly
generated with high precision (with numerical grid step $\Delta t<= 2^{-31}$), using recursive scheme 
\[
t_n=t_{n-1}+R+0.25, 
\]
where $R$ is random computer generated
part of the full time step ($0<R<1$). The obtained time points were then scaled so that $t_1=0.0$ and $t_N=N-1$ fixing mean step length
at $\bar \Delta t = 1.0$ and consequently the standard Nyquist frequency at $0.5$. In recursive  scheme the fixed part ($0.25$) of the time step
guarantees that in final sequence no step is shorter than $0.2$ (the corresponding Nyquist frequency is then $2.5$).
Formal Nyquist limit which can be computed from grid step was approximately $2^{30}$ - or in practical terms - infinity. 
The frequency
search range for all runs was fixed at $0.1-500.0$ (thousand times the normal range and two hundred times of the
range which comes from shortest time distance argument). Step size in frequency was taken as
it is normally done $\Delta s = 1/N$. Corresponding spectra contained then 2000-98980 points
(in the range of data point numbers from $N=3$ to $N=100$). For each trial frequency the standard Lomb-Scargle~(Lomb~1976, Scargle~1982)
statistic was computed. 

To evaluate and compare different spectra we used contrast ratio of the amplitude of the strongest
peak in the spectrum to the second strongest. Very often the strongest peaks
are evaluated against the mean spectrum level but here we are much more stringent and use highest accidental value from the
full spectrum. If the strongest peak in the spectrum was not at correct position the
contrast ratio was set to impossible value $0.5$.
The model oscillation frequency was taken quite high - $s = 444.444$ \footnote{The full set of tables from which
figures are compiled can be found {\it http://www.aai.ee/$\sim$pelt/soft.htm}.}.

\vbox{
\centerline{\psfig{figure=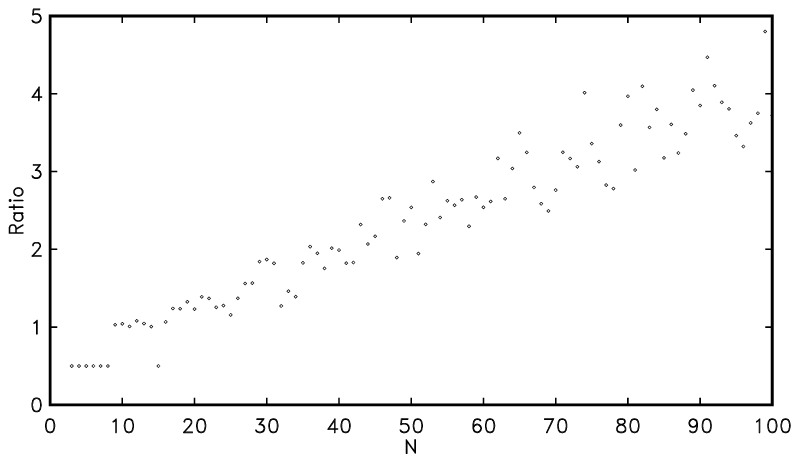,width=120truemm,angle=0,clip=}}
\vspace{-5mm}
\captionc{3}{Contrast ratios for data sets with different lengths, noiseless case.}
}
\vspace{3mm}

As seen from the Figure 3 where one concrete run for a noiseless case $E=0$ is
depicted the contrast ratio shows systematic rising trend. The inherent scatter in the contrast ratio
values is a result of random distribution of the time points. Because theoretically Lomb-Scargle spectrum
has exponential statistical distribution (see Scargle~1982) the occurrence of random strong peaks
is quite probable. Nevertheless, it can be seen that already starting from $N=17$ the correct frequency is stably
recovered and starting from $N=34$ ratio is permanently higher than $1.5$. 

The picture of the noisy case with $E=0.2$ ($10\%$ noise level ) is quite similar (see Figure 4).
The row of the correct detections starts at $N=26$ and the $1.5$ level is achieved
at nearly the same level as the noiseless case $N=43$.
\vbox{
\centerline{\psfig{figure=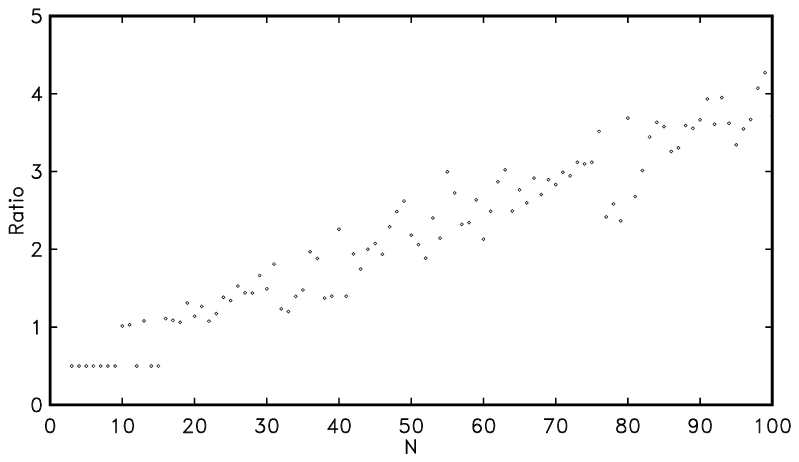,width=120truemm,angle=0,clip=}}
\vspace{-5mm}
\captionc{4}{Contrast ratios for data sets with different lengths, $10\%$ noise.}
}
\vspace{3mm}

\vbox{
\centerline{\psfig{figure=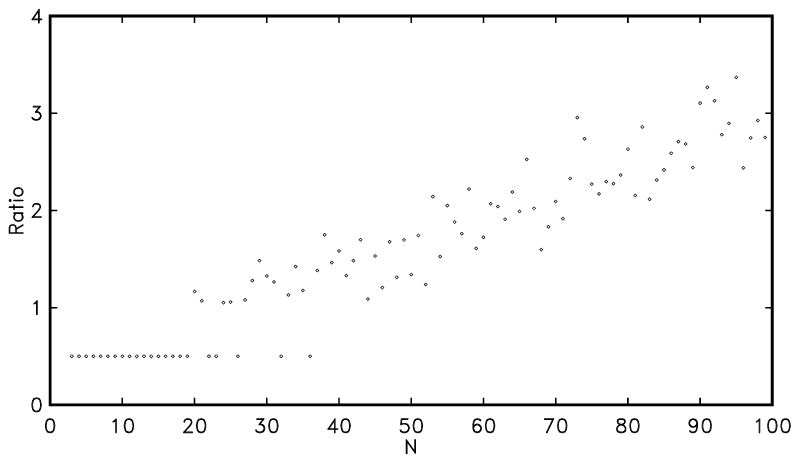,width=120truemm,angle=0,clip=}}
\vspace{-5mm}
\captionc{5}{Contrast ratios for data sets with different lengths, $25\%$ noise.}
}
\vspace{3mm}

For higher noise levels the picture starts to change. For instance if $E=0.5$ ($25\%$ noise, Figure 5) the ambiguity for
short sequences last longer and overall scatter of ratio values is higher. The monotonous exact recovery starts
at only $N=28$. The $1.5$ level stability is achieved at $N=69$. There is no need to say that these are just
particular runs and the results depend on actual random time sequences. However, the general picture is always
quite similar.

At the first glance these results seem to be extremely counter intuitive. We compute from, say $50$, observations spectra which contain
$\approx 50000$ points. In fixed time step situation we can have only $25$ statistically independent frequencies below the Nyquist frequency.
Now we claim that irregularly sampled scheme can detect true spectrum lines among $50000$ different possible positions and quite
predictably so. How this can be?

Computed values in the long spectrum for irregulrly sampled data are statistically strongly correlated. If we return to our
points on circle analogue, then  this kind of {\it deterministic} systems show also very long (essentially infinite) correlations.
Nevertheless the states where all points occur in a certain restricted position are extremely rare 
(you can compare the situation with molecules in the air).
This shows that estimation of the power spectrum
(statistical analysis) significantly differs from periodicity seeking (kind of harmonic analysis). 
In the first case we do not assume that computed spectrum has any prescribed form. Then it is quite clear that
we cannot properly estimate more than $N$ (or $N/2$ in power terms) spectrum values from $N$ data points. 
In the second case we assume that the spectrum contains only one or small number of peaks and our task is to find
the exact positions of these peaks. As we demonstrated earlier the frequency range for such analysis is 
limited by possible return of phase configurations or by the modified Nyquist limit.  

\sectionb{7}{CONCLUSIONS}
From analysis above we can make following conclusions:
\begin{itemize}
\item When Galileo used  pulse counts and pendula for timing in his experiments and the obtained
precision was certainly not better than around one tenth of the second, then modern clocking
allows time measurements whose precision can exceed $10^{ - 16}$ seconds and in the 
near future even more. This allows to widen frequency range for ``time spectrometers'' significantly. Even when concrete experiment is seeking for
phenomena which show up at lower frequencies, it is always reasonable to record
timing data with highest precision available. It is never ruled out that
certain follow-up analysis of archival data can make this information usable.
\item It is always reasonable to keep integration times with fixed and precisely measured length. For future archival work it is also important to give exact positions of published time points against integration intervals (start, exact
middle point or end). This allows properly to combine the data with future observations.
\item When analyzing already measured data it is reasonable to compute observing windows for a wide range frequencies to check for the possible peaks which occur
not only because of some inherent periodic gaps in timing points but also because of possibility of returning phase configurations. This is especially true for data
which consist of small number densely populated short fragments.
\item In some experiments the maximum time sampling frequency is limited
by technical set up (dead times, recording lags etc). It is then always
possible to widen frequency range for periodicity search by randomizing measurement
moments. If the fixed step scheme is unavoidable, even then we can search for higher frequency aliases at lower parts of spectra. For proper final frequency localization some physical insight can be applied or if possible - prefiltering of the data.
\end{itemize}
We hope that our considerations help to dispel some common misunderstandings and
point to new perspectives of high frequency range experiments which can be of
great astrophysical interest.

ACKNOWLEDGEMENTS. This work was supported by the Estonian Science Foundation
grant No. 6813. I thank T\~onu Viik for a kind assistance with manuscript preparation.

\References

\refb Bracewell~R.~N. 2000, The Fourier Transform and Its Applications (McGraw-Hill, Singapore), 219

\refb Deeming~T.~J. 1975, ApSS 36, 137

\refb Eyer~L., Bartholdi~P. 1999, A\& AS 135, 1

\refb Jurkevich~I. 1971, ApSS 13, 154

\refb Kac~M. 1947, Bull. of the Am. Math. Soc 53, 1002

\refb Koen~C. 2006, MNRAS 371, 1390

\refb Kurtz~D.~W. 1983, IBVS 2285, 1

\refb Lafler~J., Kinman~T.~D. 1965, ApJS 11, 216

\refb Lomb~N.~R. 1976, ApSS 39, 447

\refb Press~W.~H., Teukolsky~S.~A., Vetterling~W.~T., Flannery~B.~P. 2007, {\it Numerical Recipes} (Cambridge University Press) 685

\refb Scargle~J.~D. 1982, ApJ 263, 835

\refb Schwarzenberg-Czerny~A. 1989, MNRAS 241, 153

\end{document}